\documentclass[aps,prl,twocolumn,superscriptaddress,10pt]{revtex4-2}

\usepackage{siunitx}
\usepackage{amsfonts}
\usepackage{amsmath}
\usepackage{amssymb}
\usepackage{graphicx}
\usepackage{dcolumn}
\usepackage{mciteplus}
\usepackage{euscript}
\usepackage{multirow}
\usepackage{color}
\usepackage{bm}
\usepackage{wasysym}  

\usepackage{footnote}

\graphicspath{{./figs/}{../figs/}} 

\newcommand{\tc}{T_\text{C}}

\begin{document}

\title{Controlling Skyrmion Lattices via Strain: Elongation, Tilting, and Collapse Mechanisms}
\author{Haijun Zhao}
\thanks{Correspondence to: haijunzhao@seu.edu.cn}
\affiliation{School of Physics, Southeast University, 211189 Nanjing, China.}
\affiliation{Ames National Laboratory, U.S. Department of Energy, Ames, Iowa 50011, USA}
\author{Tae-Hoon Kim}
\affiliation{Ames National Laboratory, U.S. Department of Energy, Ames, Iowa 50011, USA}

\author{Lin Zhou}

\affiliation{Ames National Laboratory, U.S. Department of Energy, Ames, Iowa 50011, USA}
\affiliation{Department of Materials Science and Engineering, University of Virginia, Charlottesville, VA 22904}
\author{Liqin Ke}
\thanks{Correspondence to: liqin.ke@virginia.edu}
\affiliation{Ames National Laboratory, U.S. Department of Energy, Ames, Iowa 50011, USA}
\affiliation{Department of Materials Science and Engineering, University of Virginia, Charlottesville, VA 22904}

\date{\today}

\begin{abstract}
This study establishes a comprehensive framework for the three-dimensional strain control of magnetic skyrmion strings. We integrate analytical modeling, micromagnetic simulations, and \textit{in situ} Lorentz transmission electron microscopy experiments to demonstrate that externally applied strain is a powerful stimuli for manipulating three-dimensional magnetic skyrmion strings. Analytical models predict that strain induces both elongation and bidirectional tilting of skyrmion strings in bulk systems, a finding corroborated by numerical simulations. 
These simulations further reveal that strain
drives the system from fragmented multi-domain states toward unified single-domain configurations and facilitates skyrmion string rupture via bobber formation at critical strain levels.
The collapse of the skyrmion lattice exhibits a temperature-dependent character, shifting from first-order to second-order behavior near the critical temperature $T_c$. Reducing sample thickness significantly increases the critical strain required for annihilation due to the suppression of tilting. Experimental validation on a $\text{Co}_8\text{Zn}_{8.5}\text{Mn}_{3.5}$ sample confirms strain-induced elongation and subsequent collapse into a conical phase via anti-cluster formation, directly implicating strain-modulated Dzyaloshinskii-Moriya interaction (DMI) as the primary mechanism in this system, over magnetocrystalline anisotropy. These findings provide a mechanistic understanding of strain-mediated control in three-dimensional magnetic systems, demonstrating its feasibility for energy-efficient spintronic applications.

\end{abstract}

\maketitle

\maketitle
\section{I. introduction}
Magnetic skyrmions, nanoscale vortex-like swirling spin structures within two-dimensional (2D) systems, exhibit remarkable topological properties~\cite{ISI:A1989T174900017,muhlbauer2009Skyrmion,yu2010Realspace,kim2021Kinetics,kim2020Mechanisms,tokura2021Magnetic}.
Typically found in chiral magnets, their stability stems from the intricate interplay between exchange interactions, Zeeman effects, and the Dzyaloshinskii-Moriya interaction (DMI), which necessitates a noncentrosymmetric crystal lattice, as observed in canonical helimagnetic materials like MnSi~\cite{muhlbauer2009Skyrmion}, Fe$_{1-x}$Co$_x$Si\cite{yu2010Realspace}, FeGe\cite{yuRoomtemperatureFormationSkyrmion2011}, Cu$_2$OSeO$_3$\cite{seki2012Observationa}, and $\text{Co}_x\text{Zn}_{y}\text{Mn}_z$ ($x + y + z = 20$) alloys~\cite{kim2021Kinetics,kim2020Mechanisms}.
In stark contrast to conventional ferromagnetic domain walls, skyrmions can be manipulated with electric currents that are five orders of magnitude smaller~\cite{jonietz2010Spin,yuSkyrmionFlowRoom2012,jiangBlowingMagneticSkyrmion2015,pham2024Fast,iwasaki2013Universal_}, making them promising candidates for future spintronic technologies, including advanced information storage systems~\cite{tomaselloStrategyDesignSkyrmion2015,zhengExperimentalObservationChiral2018}, logical devices~\cite{zhang2015Magnetic_,luo2021Skyrmion,mak2022Singlebit_}, and unconventional computing paradigms~\cite{liMagneticSkyrmionsUnconventional2021,lee2022Reservoir_,pinna2020Reservoir_,prychynenko2018Magnetic_a,lee2023Handwritten_}.

In three-dimensional systems, skyrmions form string-like structures—uniform stacks of two-dimensional skyrmion spin textures~\cite{sekiDirectVisualizationThreedimensional2022,kim2020Mechanisms}—analogous to pancake vortex flux lines in type-II superconductors~\cite{Blatter1994,Zhao2016}.
These strings remain straight in a ferromagnetic background~\cite{sekiDirectVisualizationThreedimensional2022}, but become twisted when embedded in a conical magnetic phase~\cite{kim2020Mechanisms}.
When skyrmion strings terminate within the material or interconnect, they give rise to magnetic bobbers\cite{rybakovNewTypeStable2015,ranCreationChiralBobber2021,zhengExperimentalObservationChiral2018,kim2020Mechanisms,sekiDirectVisualizationThreedimensional2022} or Y-shaped skyrmion strings\cite{mildeUnwindingSkyrmionLattice2013a,kim2020Mechanisms,sekiDirectVisualizationThreedimensional2022}, respectively.
Both types of structures inherently host Bloch points, which are magnetic singularities characterized by a vanishing magnetization magnitude.

Various methods have been employed to manipulate skyrmions, including magnetic fields~\cite{mildeUnwindingSkyrmionLattice2013, kim2020Mechanisms, kim2021Kinetics}, heating~\cite{kim2021Kinetics, kong2013Dynamics, yuRoomtemperatureFormationSkyrmion2011a}, electric currents~\cite{jiangDirectObservationSkyrmion2017a, yuSkyrmionFlowRoom2012, yuCurrentInducedNucleation2017}, microwaves~\cite{weilerHelimagnonResonancesIntrinsic2017a, mochizukiSpinWaveModesTheir2012a}, and strain~\cite{shibataLargeAnisotropicDeformation2015a, zhangStrainDrivenDzyaloshinskiiMoriyaInteraction2021a, leiStraincontrolledMagneticDomain2013, kongDirectObservationTensilestraininduced2023,niiUniaxialStressControl2015,chacon2015Uniaxial_b,seki2017Stabilization_b,tanaka2020Theoretical_c}.
Among these, strain control has garnered significant interest due to its low energy consumption and versatile capabilities in modulating skyrmions, such as elongating them~\cite{shibataLargeAnisotropicDeformation2015a, dai2022Elongationa}, reversibly moving them~\cite{liuStrainInducedReversibleMotion2024}, and creating or annihilating them~\cite{zhang2021straindrivena, niiUniaxialStressControl2015}.
This strain-mediated control therefore provides an energy-efficient means for skyrmion manipulation, highlighting its application potential.
For instance, by controlling the elongation of skyrmions, the Hall angle can be dynamically adjusted~\cite{xia2020Dynamics}, while the creation and annihilation of skyrmions are pivotal steps in digital writing processes~\cite{tomaselloStrategyDesignSkyrmion2015}.
Furthermore, strain modulation extends beyond skyrmions, encompassing the manipulation of other magnetic structures. For example, strain can induce magnetic soliton lattice in chiral helimagnets ~\cite{paterson2020tensile}. Additionally, our recent study demonstrates that strain can control the reorientation of helical phase directions ~\cite{kim2025Topological_}.

To elucidate the physics underlying strain-modulated spin textures, various pivotal effects are considered. 
According to classical magnetoelastic theory,
uniaxial stress induces magnetic anisotropy~\cite{plumerMagnetoelasticEffectsSpindensity1982, niiUniaxialStressControl2015,wang2018Uniaxial_}.
When perpendicular stress is applied, varying stress mimics the effect of an external magnetic field, enabling control of skyrmion creation and annihilation~\cite{niiUniaxialStressControl2015}.
Strain may also modify the DMI, either enhancing, reducing, or inducing it along the strain direction, thereby altering spin textures. In 2D systems, in-plane strain breaks DMI isotropy, producing elliptical skyrmions~\cite{shibataLargeAnisotropicDeformation2015a, dai2022Elongationa}.
In centrosymmetric La$_{0.67}$Sr$_{0.33}$MnO$_3$, where DMI is symmetry-forbidden, strain gradients can create effective DMI, stabilizing helical phases and skyrmions~\cite{zhang2021straindrivena}. For non-centrosymmetric Mn$_{1-x}$Fe$_x$Ge, compressive stress suppresses DMI along the stress axis, changing magnetic phase stability~\cite{koretsune2015control_}. Our recent study on $\text{Co}_8\text{Zn}_{8.5}\text{Mn}_{3.5}$ demonstrates that applied strain induces perpendicular alignment of the helical spin propagation vector relative to the strain axis, a reorientation mechanism attributed to strain-modulated asymmetric DMI \cite{kim2025Topological_}.

Despite extensive 2D studies, the 3D response of skyrmion strings to strain induced anisotropic DMI modification remains unexplored.
This work aims to bridge this gap by combining  analytical modeling and micromagnetic simulations, revealing that strain induces not only in-plane elliptical deformation but also string tilting in 3D systems.
This tilt weakens interlayer skyrmion coupling, ultimately leading to string fragmentation and skyrmion annihilation.
Our theoretical predictions are experimentally validated through \emph{in-situ} Lorentz transmission electron microscopy (LTEM) observations of strain-mediated skyrmion dynamics under controlled uniaxial tensile stress.

The paper is organized as follows. In Section II, we derive the theoretical framework for strain-induced tilting of skyrmion strings from anisotropic DMI theory. Section III presents micromagnetic simulations that corroborate our analytical predictions. Section IV provides experimental validation through direct observation of strain-mediated skyrmion dynamics. Finally, we conclude with a summary of key findings  in Section V. 

\section{II. Strain-induced tilting of skyrmion strings derived from anisotropic DMI theory }
The free energy functional is expressed as \cite{landauElectrodynamics_}:
\begin{eqnarray}
F &=& F_{ex} + F_{DMI} + F_{h} + F_{a} \nonumber\\
&=& \int_{V} \left( A_{ij} \partial_{i} \mathbf{m} \cdot \partial_{j} \mathbf{m} + D_{ijk} m_{i} \partial_{j} m_{k} \right) d\tau + F_{h} + F_{a},\nonumber\\ \label{eq:model1}
\end{eqnarray}
where Einstein's summation convention is applied. Here, \(A_{ij}\) and \(D_{ijk}\) are the exchange stiffness tensor and the Dzyaloshinskii-Moriya interaction (DMI) tensor, respectively. 
$\mathbf{m}=(m_i,m_j,m_k)$ is the unit vector representing the magnetization direction,
\(F_{h}\) describes the contribution of the external magnetic field, and \(F_{a}\) represents the anisotropy energy, including dipolar interactions.

For FeGe\cite{yuRoomtemperatureFormationSkyrmion2011}, Cu$_2$OSeO$_3$\cite{seki2012Observationa}, and Co$_x$Zn$_y$Mn$_z$ ($x + y + z = 20$) alloys~\cite{kim2021Kinetics,kim2020Mechanisms}, the exchange stiffness and DMI are isotropic in the absence of strain, where \(A_{ij} = A \delta_{ij}\) and \(D_{ijk} = D \epsilon_{ijk}\), with \(\epsilon_{ijk}\) being the Levi-Civita symbol. This simplifies the energy functional to:
\begin{eqnarray}
F &=& \int_{V} \left[ A (\nabla \mathbf{m})^{2} + D_{x} L_{yz}^{(x)} + D_{y} L_{zx}^{(y)} + D_{z} L_{xy}^{(z)} \right] d\tau \nonumber \\
&&+ F_{h} + F_{a} \label{eq:model2}\\
&=& \int_{V} \left( A (\nabla \mathbf{m})^{2} + D \mathbf{m} \cdot \nabla \times \mathbf{m} \right) d\tau + F_{h} + F_{a}. \label{eq:model3}
\end{eqnarray}
 Here, 
\[
L_{jk}^{(i)} = m_{k} \frac{\partial}{\partial i} m_{j} - m_{j} \frac{\partial}{\partial i} m_{k}
\]
represents the Lifshitz invariant.
Equation (\ref{eq:model3}) constitutes the widely adopted standard model \cite{dzyaloshinskii1965Theory_, leonov2016Chiral_, hals2017New_, rybakov2016New_}.
When strain is applied, the isotropic DMI tensor becomes anisotropic, causing  the form of  Eq. (\ref{eq:model3}) to revert to that of Eq. (\ref{eq:model2}). A strong field is applied to stabilize the skyrmion lattice, making the Zeeman energy dominate the magnetic anisotropy (which includes dipolar interactions); we therefore neglect $F_a$ to simplify the model.
Without loss of generality, we assume that strain is applied in the \(\hat{x}\) direction. Following Ref. \cite{shibata2015Large_,kim2025Topological_}, the dominant effect of the applied strain is the adjustment of the DMI strength along the strain direction, namely, 
\begin{equation}
D_{x}(\sigma) = [1 - \eta(\sigma)]D. \label{eq:DX}
\end{equation}
Here, \(\eta\) acts as an effective strain parameter that increases from 0 to 1 as the strain \(\sigma\) increases. \(D_{y}\) and \(D_{z}\) remain unaffected, i.e., 
\begin{equation}
D_{y} = D_{z} = D_{x}(0) = D. \label{eq:DY}
\end{equation} 

To investigate the tilting behavior of the skyrmion string, we start from 2D model and consider the 3D tilting as a perturbation.

The magnetic configuration of a 2D skyrmion crystal (SkX) state can be written as:
\begin{equation}
\mathbf{m} = m_{0} \mathbf{e}_{z} + \sum_{\alpha=1}^{3} \left( \mathbf{m}_{\alpha} e^{i \mathbf{Q}_{\alpha} \cdot \mathbf{r}} + \mathbf{m}^{*}_{\alpha} e^{-i \mathbf{Q}_{\alpha} \cdot \mathbf{r}} \right). \label{eq:m_sk}
\end{equation}
The three vectors \(\mathbf{Q}_{i}\) (\(i = 1, 2, 3\)) are magnetic modulation vectors in the \(xy\)-plane and satisfy the constraint for SkX formation:
\begin{equation}
\sum_{\alpha=1}^{3} \mathbf{Q}_{\alpha} = 0.
\end{equation}
We now retain the freedom along the $\hat{z}$ direction while assuming the skyrmion string is displaced along the $\hat{x}$ direction (i.e., the strain direction). In this scenario, interlayer sliding is treated as a perturbation. 
By neglecting other variations, we can express the derivative along $\hat{z}$ as
\[
\frac{\partial}{\partial z} = \frac{\partial}{\partial x} \frac{\partial x}{\partial z} = k \frac{\partial}{\partial x}.
\]
It should be noted, however, that this approximation breaks down near the sample surfaces (within a depth of $< 0.5L_D$, where $L_D = 4\pi A / D$ is the DMI length), where surface twist effects become significant. As a skyrmion approaches the surface, its helicity shifts from the Bloch-type configuration observed in the midsection and acquires a partially Néel-type character \cite{rybakov2013Threedimensional_, leonov2016Chiral_, zhang2018direct_}. This helicity gradient inhibits full overlap of skyrmions in adjacent layers under the present approach. Therefore, our model remains valid only within the midsection of thick samples (with thickness $L_z > L_D$). The dynamics in thin samples, as well as the surface effects in thick samples, are fully captured in our following numerical simulations, and our primary analysis focuses on the tilting and rupture mechanisms.

Substituting Eq. (\ref{eq:m_sk}) to Eq. (\ref{eq:model2}), and omitting the constant terms, 
the free energy now becomes:
\begin{equation}
F = \sum_{\alpha=1}^{3} \left[ \mathbf{m}_{\alpha}^{*} \right]^{T} M_{\alpha} \mathbf{m}_{\alpha},
\label{eq:model4}
\end{equation}
where \(M_{\alpha}\) is defined as:
\begin{equation}
M_{\alpha} = \begin{pmatrix}
E_{\alpha}^{ex} & 2i D_z k Q_{x,\alpha} & -2i D_y Q_{y,\alpha} \\
-2i D_z k Q_{x,\alpha} & E_{\alpha}^{ex} & 2i D_x Q_{x,\alpha} \\
2i D_y Q_{y,\alpha} & -2i D_x Q_{x,\alpha} & E_{\alpha}^{ex}
\end{pmatrix},
\label{eq:M_alpha}
\end{equation}
with 
\begin{equation}
    E_{\alpha}^{ex} = A \left( Q_{x,\alpha}^2 + Q_{y,\alpha}^2 + k^2 Q_{x,\alpha}^2 \right).\label{eq:E_alpha_ex}
\end{equation}

The lowest eigenvalue is:
\begin{equation}
E_\alpha = E_{\alpha}^{ex} - 2 \sqrt{D_x^2 Q_{x,\alpha}^2 + k^2 D_z^2 Q_{x,\alpha}^2 + D_y^2 Q_{y,\alpha}^2}. \label{eq:E_alpha_lowest}
\end{equation}
For \(k = 0\), Eqs. (\ref{eq:model4}-\ref{eq:E_alpha_lowest})  reduce to the 2D case. Minimization of the free energy yields the three wave vectors of the skyrmion lattice state \cite{shibata2015Large_}:
\begin{equation}
\mathbf{Q}_{1} = \left(0, \frac{D_{y}}{A}\right), \label{eq:Q1}
\end{equation}
\begin{equation}
\mathbf{Q}_{2} = \left(\frac{D_{x}}{A}\sqrt{1 - \frac{D_{y}^{4}}{4 D_{x}^{4}}}, -\frac{D_{y}}{2A}\right), \label{eq:Q2}
\end{equation}
\begin{equation}
\mathbf{Q}_{3} = \left(-\frac{D_{x}}{A}\sqrt{1 - \frac{D_{y}^{4}}{4 D_{x}^{4}}}, -\frac{D_{y}}{2A}\right). \label{eq:Q3}
\end{equation}
We now assume these expressions remain valid in the 3D case, provided tilting can be treated as a perturbation.  For \(\mathbf{Q}_{1}\), \(Q_{x,\alpha=1} = 0\), resulting in \(E_{\alpha=1}\) being independent of \(k\) [Eqs. (\ref{eq:E_alpha_ex}-\ref{eq:E_alpha_lowest})]. This result is expected since \(Q_x = 0\) implies the propagation vector lies exclusively along the y-axis. Consequently, the system possesses perfect translational symmetry along the x-direction, which makes sliding or tilting along $x$ physically meaningless.

\begin{figure}[tbp]
    \centering
    \includegraphics[width=0.98\linewidth]{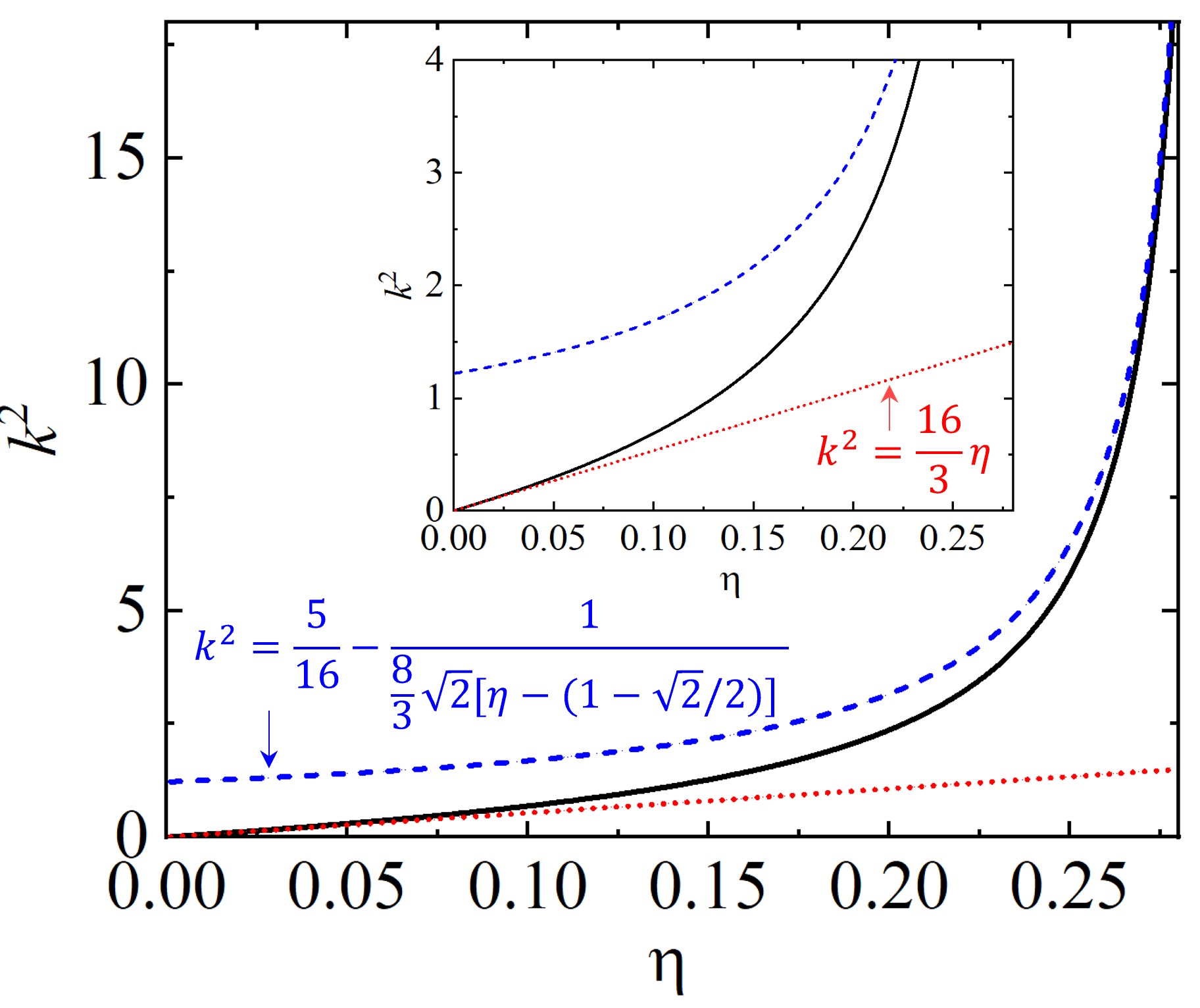} 
    \caption{Analytically derived relationship between the square of slope $k^2$ and the effective strain $\eta$ (solid black line). This relationship demonstrates linearity for small values of $\eta$ (dotted red line), yet exhibits a transition to inverse behavior as $\eta$ approaches the singular point $\eta=1-\sqrt{2}/2\approx0.3$ (dashed blue line). The \textbf{inset} provides an enlarged view of the small-$k^2$ region, which highlights the linear regime.}
    \label{fig:k2_vs_eta}
  \end{figure}

For both \(\mathbf{Q}_{2}\) and \(\mathbf{Q}_{3}\) [Eqs. (\ref{eq:Q2}) and (\ref{eq:Q3})],
\begin{equation}
Q_x^2 = \frac{D_x^2}{A^2} - \frac{D_y^4}{4 A^2 D_x^2}, \label{eq:Qx2}
\end{equation}
and
\begin{equation}
Q_y^2 = \frac{D_y^2}{4 A^2}. \label{eq:Qy2}
\end{equation}

Substituting Eqs. (\ref{eq:Qx2}) and (\ref{eq:Qy2}) into Eq. (\ref{eq:E_alpha_lowest}) and minimizing \(E_\alpha\) with respect to $k^2$, we obtain:
\begin{eqnarray}
&&k^{2} = \dfrac{4\eta(2-\eta)(1-\eta)^{2} ( \eta^2-2\eta+2)}{4(1-\eta)^{4} - 1}\nonumber \\
&&\approx 
\begin{cases}
 \frac{16}{3}\eta & \text{for } \eta \thicksim 0, \\
 \frac{5}{16} - \frac{1}{\frac{8}{3} \sqrt{2} \left[\eta - (1-\sqrt{2}/2)\right]} & \text{for } \eta \thicksim 1-\sqrt{2}/2.  
\end{cases}
\label{eq:k2}
\end{eqnarray}

The quadratic dependence of $k^2$ (rather than $k$)
on $\eta$ reflects the fact that the sign of the tilt is physically irrelevant. 
This $\mathbb{Z}_2$ symmetry is a  consequence of $C_2$ rotational invariance (180\textdegree{} about $\hat{z}$): 
\begin{itemize}
    \item  The core energy contributions (exchange, DMI, and Zeeman coupling to the $\hat{z}$-aligned field) are intrinsically rotationally symmetric [Eq. (\ref{eq:model2})]
    \item Uniaxial strain along $\hat{x}$ preserves this $C_2$ symmetry via counter-directed forces
    \item  $C_2$ rotation transforms $+k \rightleftarrows -k$
\end{itemize}
Consequently,  states with $\pm k$ are energetically degenerate.

As shown in Fig.~\ref{fig:k2_vs_eta}, for small $\eta$, $k^2$ varies linearly until transitioning to inverse behavior at $\eta_c = 1 - \sqrt{2}/2 \approx 0.3$, where divergence signals string rupture and lattice collapse.

\section{III. Micromagnetic Simulation }
\subsection{Methodology}
The micromagnetic simulations are performed by numerically solving the Landau-Lifshitz-Gilbert (LLG) equation:

\begin{equation}
\frac{1 + \alpha^{2}}{\gamma} \frac{\partial \mathbf{m}}{\partial t} = -\mathbf{m} \times \mathbf{H}_{\text{eff}} - \alpha \mathbf{m} \times \left( \mathbf{m} \times \mathbf{H}_{\text{eff}} \right),
\end{equation}

where \(\alpha\) and \(\gamma\) are the dimensionless Gilbert damping parameter and the gyromagnetic ratio, respectively, and \(\mathbf{H}_{\text{eff}}\) is the effective magnetic field, defined as:

\begin{equation}
\mathbf{H}_{\text{eff}} = -\frac{1}{\mu_0 M_s} \frac{\delta F}{\delta \mathbf{m}} + \mathbf{H}_{\text{thermal}}.
\end{equation}

Here, \(F\) is the total free energy functional given by Eq. (\ref{eq:model1}), \(M_s\) is the saturation magnetization, and \(\mu_0\) is the vacuum permeability. The thermal field \(\mathbf{H}_{\text{thermal}}\), which accounts for stochastic thermal fluctuations, satisfies the following conditions \cite{brown1963Thermal_, leliaert2017Adaptively_}:

\begin{equation}
\langle \mathbf{H}_{\text{thermal}} \rangle = 0,
\end{equation}

\begin{equation}
\langle \mathbf{H}_{\text{thermal},i}(t) \mathbf{H}_{\text{thermal},j}(t') \rangle = \frac{2 k_B T \alpha}{\mu_0 M_s \gamma V} \delta(t - t') \delta_{ij},
\end{equation}
where \(k_B\) is the Boltzmann constant, \(T\) is the temperature, and \(V\) is the volume of the discretization cell. The thermal field can be expressed as:

\begin{eqnarray}
\mathbf{H}_{\text{thermal}}(T) &=& \boldsymbol{\xi} \sqrt{\frac{2 k_B T \alpha}{\mu_0 M_s \gamma V \Delta \tau}}\nonumber \\
&=&\boldsymbol{\xi}\sqrt{\frac{T/\tc}{\Delta\tau}}\sqrt{\frac{2k_B\tc\alpha}{M_s\gamma V}}
\nonumber \\
&=& \boldsymbol{\xi}\frac{\sqrt t}{\sqrt{\Delta\tau}}B_\text{C}.
\end{eqnarray}

where \(\boldsymbol{\xi}\) is a random vector sampled from a standard normal distribution. $\tc$ is the critical temperature above which the skyrmion lattice phase is destroyed. $V$ is the volume of the unit cell and $\Delta\tau$ is the time step. To determine $B_\text{C}$, we initially set $t=T/\tc=1$, and incrementally increase $B_\text{C}$ from zero until the system is disrupted. Subsequently, we maintain $B_\text{C}$, and adjust the thermal field by varying the effective temperature $t$ from 0 to 1.

\subsection{Simulation Setup}
The LLG equation is integrated using a self-developed GPU-accelerated code. We employ the sixth-order Runge-Kutta-Fehlberg (RKF56) method for time integration, ensuring high accuracy and stability. To be
consistent with our analytical results, the simulations are performed in dimensionless units, where:
\begin{itemize}
    \item Energy density is scaled by \(D^2 / (4A)\),
    \item Length is scaled by the DMI length \(L_D = 4\pi A / D\),
    \item Magnetic field is scaled by the saturation field \(H_D = D^2 / (2A\mu_0 M_s)\),
    \item Time is scaled by $1/(\mu_0 M_s \gamma)$.
\end{itemize}
For computational convenience, we set $A = 1$, $D = 4\pi$, and $1/\gamma=\mu_0 M_s=8\pi^2$, yielding a DMI length $L_D = 1$, the saturation field \(H_D =1\), and the characteristic time $\tau_c=1$.  
The simulation domain with dimensions $26.25L_D \times 26.25L_D \times L_z$ is discretized using a uniform mesh $\Delta x = \Delta y = \Delta z = 3L_D/40 \approx 0.075L_D$. 
Validation studies with finer meshes ($\Delta x = \Delta y = L_D/16$, $\Delta z = L_D/32$) confirm negligible numerical errors from the finite-difference discretization.
Periodic boundary conditions are applied in the $\hat{x}$ and $\hat{y}$ directions, while open boundaries in $\hat{z}$ model an infinite thin film geometry with thickness $L_z$. 

The initial defect-free skyrmion lattice phase, prior to the application of strain, is obtained by simulating the thermal nucleation process starting from a metastable cone state. Thermal fluctuations are introduced at a reduced temperature \( t = 0.6 \) using an adaptive RKF56 solver with a relative tolerance of \( 10^{-4} \), allowing the system to evolve toward its ground state. The magnetic configuration is saved every 1000 integration steps for subsequent analysis of the thermal nucleation and growth dynamics (to be reported in a forthcoming publication). The simulation is manually terminated once a stable, defect-free skyrmion lattice is visually confirmed, 
which served as the initial state for our strain-dependent simulations.
 It is important to note that the initial lattice orientation is arbitrary relative to the applied strain axis, reflecting the experimental situation where no pre-alignment is imposed.

Starting directly from such a stable skyrmion lattice, we set $t$ to specific values and apply mechanical strain by incrementally increasing the parameter $\eta$ from 0 to 1 in 100 linear steps.
At each value of $\eta$, we evolve the system for 5000 RKF56 steps to approach a steady state. This corresponds to a physical time of approximately $ 0.35/\left(\mu_0 M_s\gamma\right)$ --- $\sim1$ ns for a saturation magnetization $M_s \approx 10^5 A/m$. Convergence to a steady state was confirmed by monitoring key physical quantities ---such as energy, torque, and topological number ---
which exhibited no significant drift or only very slow, monotonic evolution
(e.g., gradual energy decrease). In most cases, the steady state was achieved within the first few hundred steps.

\begin{figure}[tbp]
  \centering \includegraphics[width=0.98\linewidth]{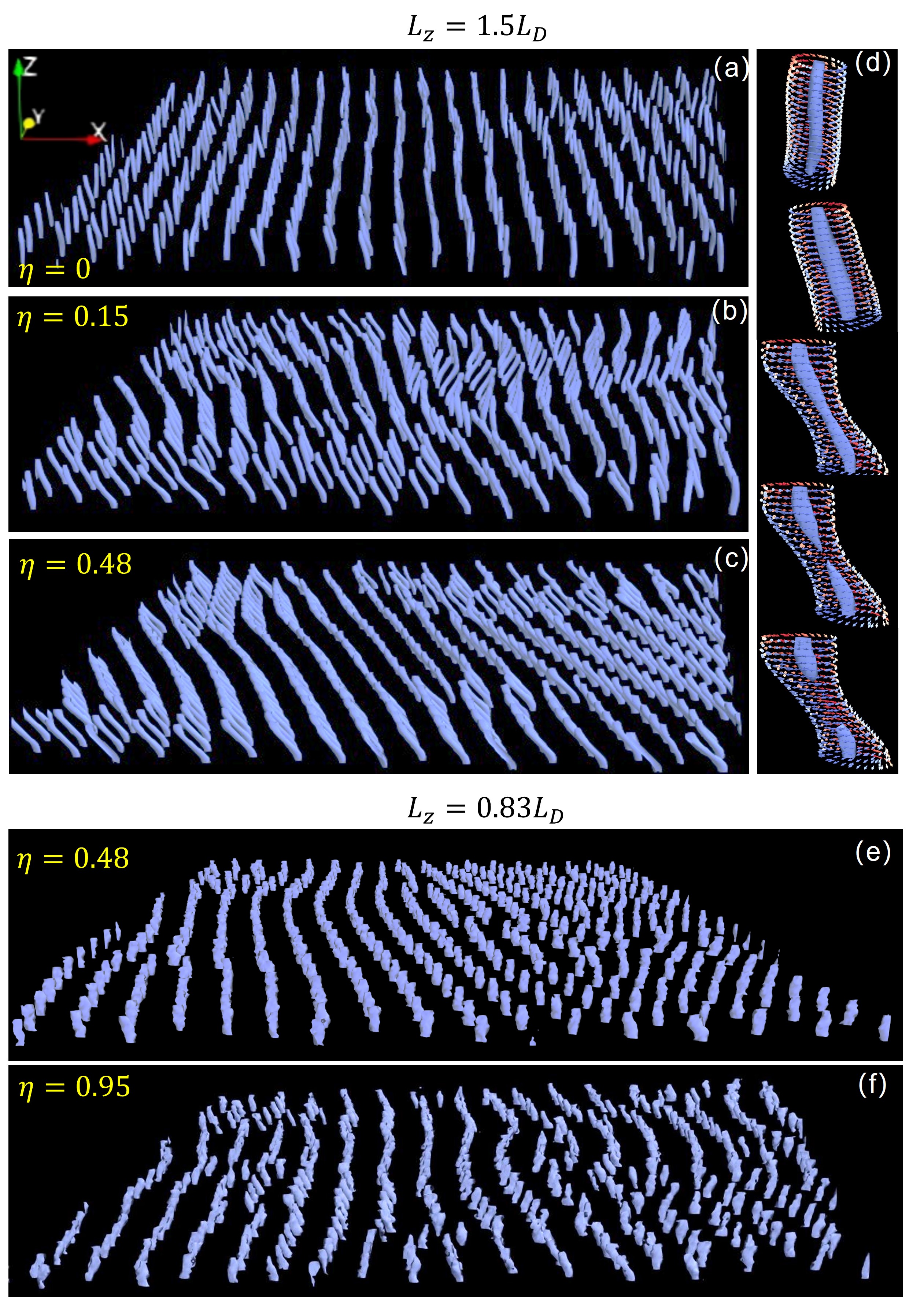} 
  \caption{Three-dimensional micromagnetic simulations of skyrmion string dynamics under uniaxial stress at $T=0.02\tc$ and $H=0.55H_D$. The thickness of the sample $L_z=1.5L_D$ for (a--d), $L_z=0.83L_D$ for (e--f). 
  Skyrmion cores (where $m_z < -0.9$) are rendered as tubes.  
  (a--c) Strain-dependent 3D configurations in a thick sample ($L_z>L_D$):  
  (a) $\eta=0$ (vertical alignment),  
  (b) $\eta=0.15$ (opposing-direction tilting with domain formation),  
  (c) $\eta=0.48$ (merged uniform tilt).  
  A preference for opposing tilt directions drives Ising-like domain formation
 in (a--b), 
 which transitions to a uniformly ordered tilt at $\eta=0.48$ in (c).  
  (d) Individual string evolution: initial tilt $\rightarrow$ rupture $\rightarrow$ surface annihilation; arrows indicate in-plane magnetization directions in the shell region ($m_z=0$).
  (e--f) Strain-dependent 3D configurations in a thin sample ($L_z<L_D$):  
  (e) $\eta=0.48$ [tilting is significantly  suppressed compared to (c)],
   (f) $\eta=0.95$ (weak tilting remains visible).
  }
  \label{fig_MMS4}
\end{figure}

\begin{figure*}[tbhp]
  \centering
  \includegraphics[width=0.98\linewidth]{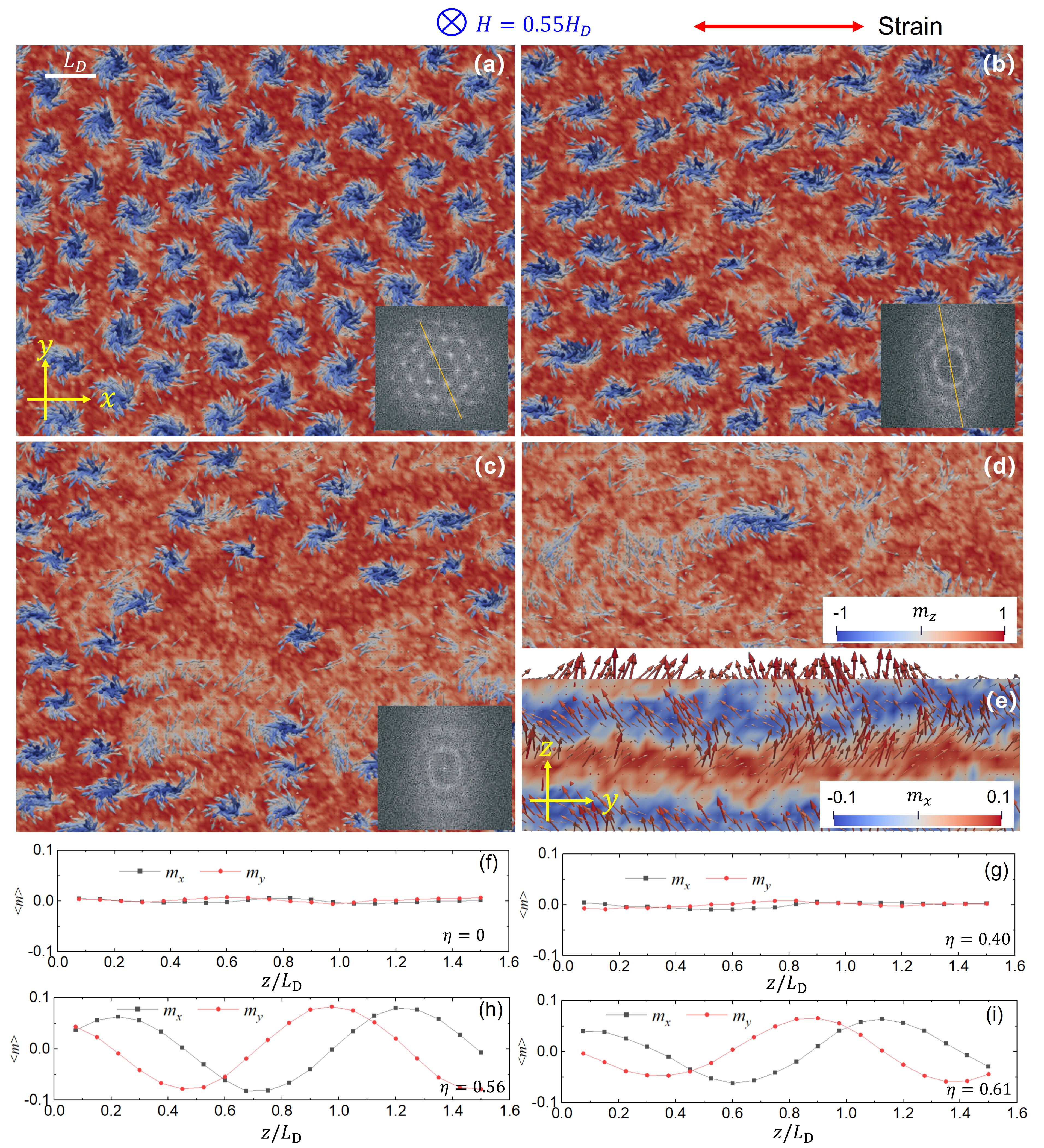} 
\caption{Evolution of the in-plane magnetic structure under uniaxial stress from micromagnetic simulations, conducted at temperature $T = 0.6T_c$, magnetic field $H = 0.55H_D$, and sample thickness $L_z = 1.5L_D$. Strain values are $\eta = 0$ (a), $\eta = 0.40$ (b), $\eta = 0.56$ (c), and $\eta = 0.61$ (d) and (e). Panels (a)--(d) show cross-sectional views in the $x$-$y$ plane at the midplane ($z = L_z/2$), viewed along $z$. Panel (e) shows the $y$-$z$ cross-section viewed along $-x$. 
(a)-(d) are colored by the $z$-component of magnetization $m_z$; in (e), color represents the $x$-component $m_x$. Arrows indicate spin orientation. Insets in (a)--(c) show FFT analyses of the corresponding skyrmion lattice structures; yellow lines mark the two crystallographic directions closest to perpendicular to the strain.
Panels (f)--(i) show the $z$-dependence of the in-plane average magnetization components $\langle m_x \rangle$ and $\langle m_y \rangle$, corresponding to the structures in (a)--(d), respectively. The sinusoidal magnetization profiles in (h) and (i), consistent with the spin structure in (g), confirm the formation of a conical phase.}
\label{fig_mms_inplane}
\end{figure*}

\subsection{Simulation Results}
The tilting of skyrmion strings is a continuous and topologically neutral deformation process. In contrast to rupture events that alter topology, the tilting itself is robust against thermal fluctuations. Although thermal agitation smears the string profiles and promotes rupture under high strain. To isolate the intrinsic mechanism without thermal blurring, we perform near-zero temperature simulations ($T = 0.02 T_c$), as shown in Fig.~\ref{fig_MMS4}.

In a thick sample ($L_z = 1.5L_D > L_D$), the tilt parameter $k \equiv \tan\theta$ (where $\theta$ is the angle between the skyrmion string and the $\hat{z}$-axis) approaches zero in the strain-free state. The slight residual tilts in Fig.~\ref{fig_MMS4}(a) arise from incomplete relaxation due to the finite-time cooling protocol, as well as from inhomogeneous microstructures near free surfaces that perturb the internal strings, inducing minor tilting and fluctuations. Notably, these residual tilts are small in magnitude, randomly oriented, and significantly weaker than those induced by thermal fluctuations at finite temperature.

Applying mechanical strain $\eta$ along a designated axis breaks rotational symmetry, inducing finite values of $k$ and leading to the formation of multi-domain configurations with alternating tilt signs [Fig.~\ref{fig_MMS4}(b)]. Many strings exhibit ``S-shaped'' profiles [e.g. the third string in Fig.~\ref{fig_MMS4}(d)], consisting of tilted midsections and straight segments near the surfaces.

These numerical results are consistent with our analytical model: the binary choice of $k$ is consistent with the $k^2$-dependence of the energy [Eq.~\eqref{eq:k2}], and the breakdown of the sliding approximation accounts for the absence of tilting near the surfaces.

We further observe numerically that multi-domain configurations with alternating $k$-signs evolve into single-domain states as the strain increases. This behavior can be understood through an analogy with domain formation in Ising systems:
\begin{itemize}
\item Under weak strain ($\eta < \eta_c$), metastable multi-domain configurations with alternating $k$-signs persist [Fig.~\ref{fig_MMS4}(b)].
\item Increasing $\eta$ enhances $|k|$ [Eq.~\eqref{eq:k2}], amplifying the structural mismatch between oppositely tilted strings.
\item The resulting increase in domain wall energy, due to this structural incompatibility, drives the system toward single-domain ordering [Fig.~\ref{fig_MMS4}(c)].
\end{itemize}
Unlike rigid Ising spins, however, skyrmion strings dynamically adjust their positions and undergo in-plane elongation during this process (see below).

A further increase in strain raises $k$, which weakens inter-layer coupling and triggers string rupture within the midsections.
This leads to the nucleation of bobber pairs [Fig.~\ref{fig_MMS4}(d)], which subsequently shrink and annihilate at the surfaces, resulting in the annihilation of individual skyrmion strings.

In contrast to thick samples, tilting is strongly suppressed in thin samples
 ($L_z < L_D$) due to the surface-twist effect. The tilt observed in a thin sample ($L_z = 0.83L_D $) under strain $\eta = 0.48$ [Fig.~\ref{fig_MMS4}(e)] is significantly smaller than that in the thick sample ($L_z = 1.5L_D $ ) under the same strain [Fig.~\ref{fig_MMS4}(c)]. Even under a substantially larger strain ($\eta = 0.95$), only weak tilting remains visible.

\begin{figure}[tbp]
    \centering
    \includegraphics[width=0.98\linewidth]{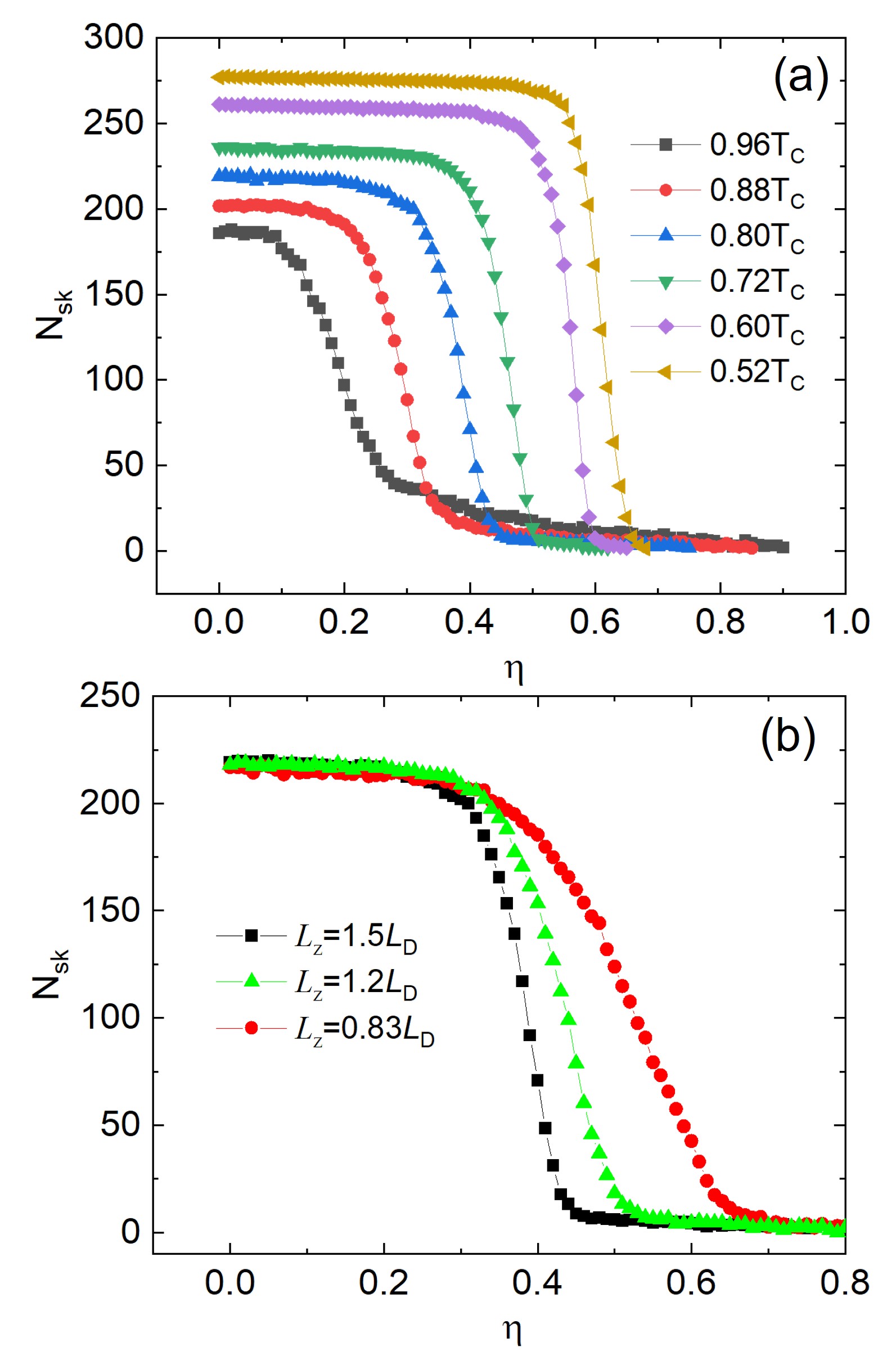} 
    \caption{
      The topological number $N_{sk}$(serves as an indicator of the skyrmion count) as a function of effective strain $\eta$  across various temperatures.
      A rapid decrease in $N_{sk}$ marks the onset of skyrmion annihilation. Notably, at lower temperatures (e.g., $T=0.52\tc$), this transition is sharp and abrupt, resembling a first-order transition.  In contrast, at higher temperatures (e.g., $T=0.96\tc$), the transition becomes smoother, characteristic of a second-order-like phenomenon.
    }
    \label{fig:slope}
  \end{figure}

Figure~\ref{fig_mms_inplane} presents the evolution of the in-plane skyrmion lattice structure in the mid-section layer under increasing strain, simulated at $T = 0.6T_c$ and $H = 0.55H_D$ for a sample thickness of $L_z = 1.5L_D$.

In the absence of strain, skyrmions maintain nearly circular shapes aside from minor thermal fluctuations [Fig.~\ref{fig_mms_inplane}(a)]. Under applied strain, two distinct responses are observed: (1) a fast process involving elongation of individual skyrmions along the strain direction, and (2) a slow, collective reorientation of the lattice, manifested by the rotation of one wave vector (indicated by the yellow line in the FFT inset) toward the direction perpendicular to the strain [Fig.~\ref{fig_mms_inplane}(b)]. This behavior resembles the strain-induced reorientation previously observed in helical structures \cite{kim2025Topological_}, where strain similarly favors alignment of the propagation vector perpendicular to the applied direction\footnote{
The small energy difference involved leads to slow reorientation kinetics, often preventing the system from reaching a true steady state within the simulation timeframe. However, simulations initialized from a pre-oriented lattice show negligible differences in fast single-skyrmion processes—such as elongation and rupture—indicating that this incomplete relaxation does not critically affect the main conclusions.
}.
A slight misalignment often persists even under full strain in both systems, attributable to the small energy difference between slightly misaligned and perfectly aligned states. These results from our 3D simulations are consistent with both 2D theoretical predictions and experimental observations in FeGe crystals~\cite{shibataLargeAnisotropicDeformation2015a}.

At larger strain values $\eta$, a novel annihilation mechanism emerges [Figs.~\ref{fig_mms_inplane}(b–e)], which has not been reported in previous 2D studies or experiments. When $\eta$ exceeds a critical threshold, annihilation begins with the formation of a point defect (a single skyrmion vacancy) [Fig.~\ref{fig_mms_inplane}(b)]. This defect expands into anti-cluster domains [Fig.~\ref{fig_mms_inplane}(c)], which eventually dominate the system and isolate remaining skyrmion clusters prior to their complete erasure [Fig.~\ref{fig_mms_inplane}(d)].
The $x$-$z$ plane spin textures in Fig.~\ref{fig_mms_inplane}(e) confirm the formation of a conical phase—rather than a ferromagnetic state—in the fully strained condition. This is further corroborated by calculations of the in-plane average magnetization components $\langle m_x\rangle$ and $\langle m_y\rangle$ along $z$: both approach zero before annihilation [Figs.~\ref{fig_mms_inplane}(f) and (g)], but evolve into sinusoidal functions with a $-\pi/2$ phase shift afterward [Figs.~\ref{fig_mms_inplane}(h) and (i)]. The emergence of these sinusoidal magnetization profiles signifies the reestablishment of a conical phase.

In contrast, skyrmion annihilation driven by an increasing perpendicular magnetic field or by strain-induced magnetic anisotropy~\cite{niiUniaxialStressControl2015} proceeds via gradual thinning of the skyrmion distribution. This difference stems from distinct inter-skyrmion interactions. In our 3D simulations, skyrmions embedded in the conical phase experience competing short-range interactions arising from exchange–DMI competition, resulting in a Lennard-Jones-like potential with a minimum at $r \approx L_D$ and a range of $\sim\!2L_D$~\cite{kim2020Mechanisms}. This potential mediates density-dependent transitions among clusters, anti-clusters, and lattice states~\cite{Zhao2012a}. In contrast, when an external field or anisotropy suppresses the conical background in favor of a ferromagnetic state, the inter-skyrmion interaction becomes purely repulsive and decays monotonically with distance~\cite{kameda2021Controllable_}. Skyrmions then gradually separate until spacings exceed $\sim\!2L_D$, behaving as isolated particles. This key distinction offers a diagnostic means to differentiate strain-modulated DMI from strain-induced magnetic anisotropy.

This lattice collapse process is quantified by the topological number:
\begin{equation}
N_{\text{sk}} = \frac{1}{4\pi}\iint \mathbf{m} \cdot \left( \frac{\partial \mathbf{m}}{\partial x} \times \frac{\partial \mathbf{m}}{\partial y} \right)  \mathrm{d}x  \mathrm{d}y,
\end{equation}
which serves as an indicator of the skyrmion count.

\begin{figure*}[tbp]
    \centering
    \includegraphics[width=0.98\linewidth]{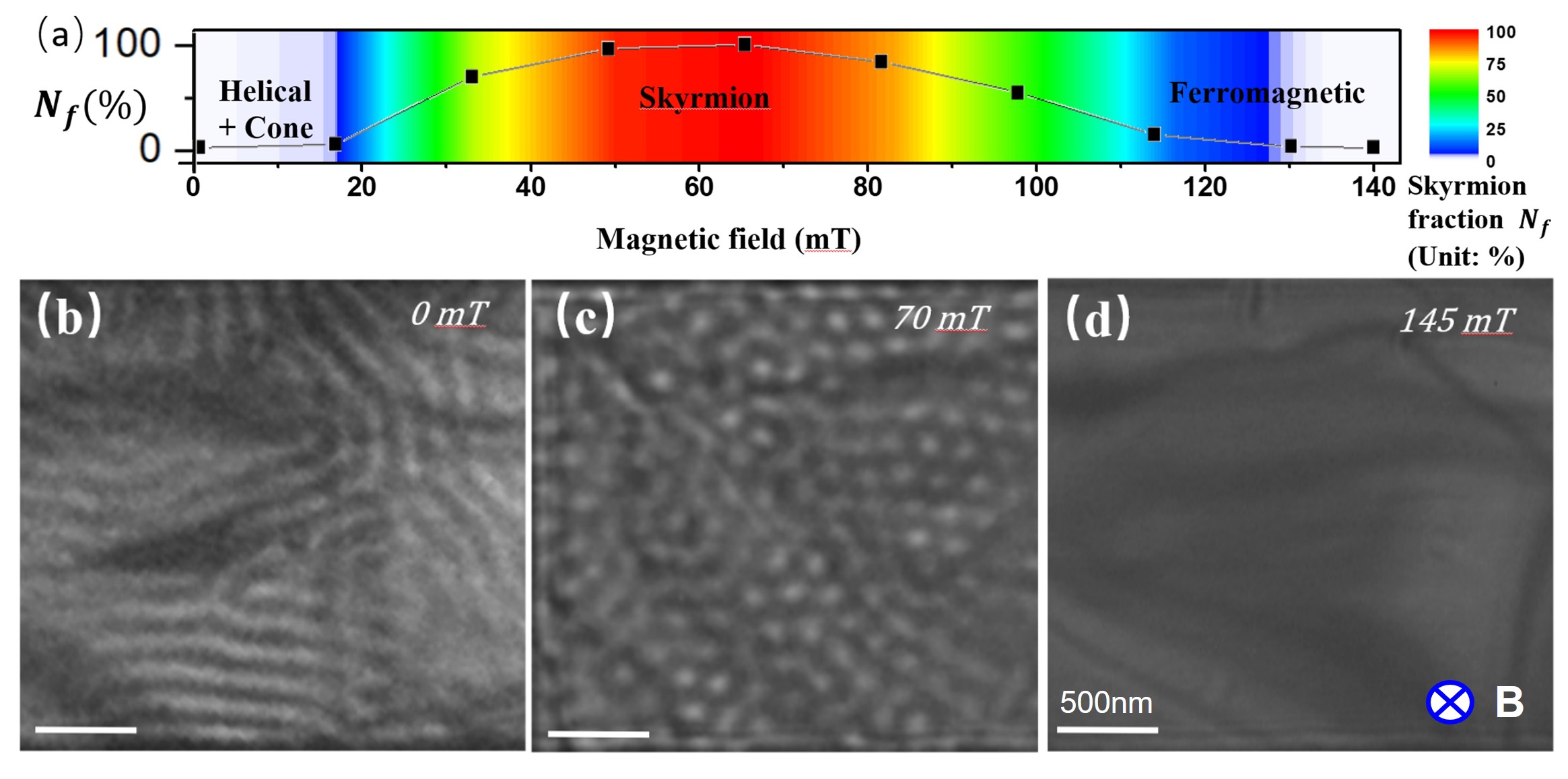} 
    \caption{ Magnetic structures and phase diagram in a $\sim$190-nm-thick {\bf(001)} $\text{Co}_8\text{Zn}_{8.5}\text{Mn}_{3.5}$ thin plate with applied strain along the [110] direction at room temperature (293 K). (a) Skyrmion fraction $N_f$ as a function of magnetic field, defined as $N_f = N(B)/N_\text{max} \times 100\%$, where $N(B)$ is the number of skyrmions at field $B$ and $N_\text{max}=N$($B$=70mT) is the maximum observed number. Both the vertical position and color represent $N_f$. (b) Underfocused ($-2$ mm) LTEM image of randomly oriented helical spin structure at zero field. (c) Skyrmion lattice structure at 70 mT. (d) Saturated ferromagnetic phase with no contrast at 145 mT.}
    \label{fig_EX1}
  \end{figure*}

Figure~\ref{fig:slope}(a) shows the averaged number of skyrmions \( N_{\text{sk}} \) (averaged along \( z \)) as a function of strain \( \eta \) for a thick sample with \( L_z = 1.5L_D \) at various temperatures. 
The topological number remains nearly constant at low strain but drops sharply to zero once $\eta$ exceeds a critical value $\eta_c$.
 This critical strain decreases with increasing temperature. The transition evolves from first-order---characterized by a discontinuous jump at low \( T \)---to second-order, exhibiting continuous variation near \( T_c \). Notably, at \( T = 0.88T_c \), the value \( \eta_c \approx 0.3 \) agrees with the analytical prediction from Eq.~(\ref{eq:k2}).

At higher temperatures, 
the reduction in the critical strain $\eta_c$ inhibits the development of long-range tilt order. The lattice collapses so rapidly that \( \pm k \)-tilted strings annihilate before the coarsening process can establish uniform tilt orientation. 
Consequently, the kinetic limitation prevents the formation of uniformly tilted domains during high-temperature collapse.

In Fig.~\ref{fig:slope}(b), we plot \( N_{\text{sk}} \) as a function of strain \( \eta \) for \( T = 0.8T_c \) across samples of varying thickness, from thick (\( L_z > L_D \)) to thin (\( L_z < L_D \)). We define two critical strains: one for the onset of the skyrmion-to-cone transition, denoted \( \eta_c^{\text{onset}} \), and another for its completion, denoted \( \eta_c^{\text{finish}} \). 
Remarkably, \( \eta_c^{\text{onset}} \) is nearly independent of sample thickness, whereas \( \eta_c^{\text{finish}} \) decreases with increasing \( L_z \), asymptotically approaching \( \eta_c^{\text{onset}} \) for \( L_z \gtrsim L_D \).

This can be understood by noting that: the onset of the transition is triggered by skyrmion elongation under increasing strain, which increases their effective size and leads to spatial crowding. The resulting in-plane repulsive interaction forces the annihilation of some skyrmions to free up space. Thus, \( \eta_c^{\text{onset}} \) is largely independent of sample thickness, as it is governed  primarily by in-plane elongation. In contrast, the completion of the transition is less influenced by in-plane skyrmion interactions due to the reduced skyrmion density, and is instead driven by the rupture of individual skyrmion strings. Tilting reduces inter-layer coupling, facilitating rupture. As discussed previously, a decrease in thickness suppresses tilting due to the surface-twist effect, thereby hindering string rupture and significantly increasing \( \eta_c^{\text{finish}} \) for \( L_z < L_D \).

  \begin{figure*}[tbp]
    \centering
    \includegraphics[width=0.98\linewidth]{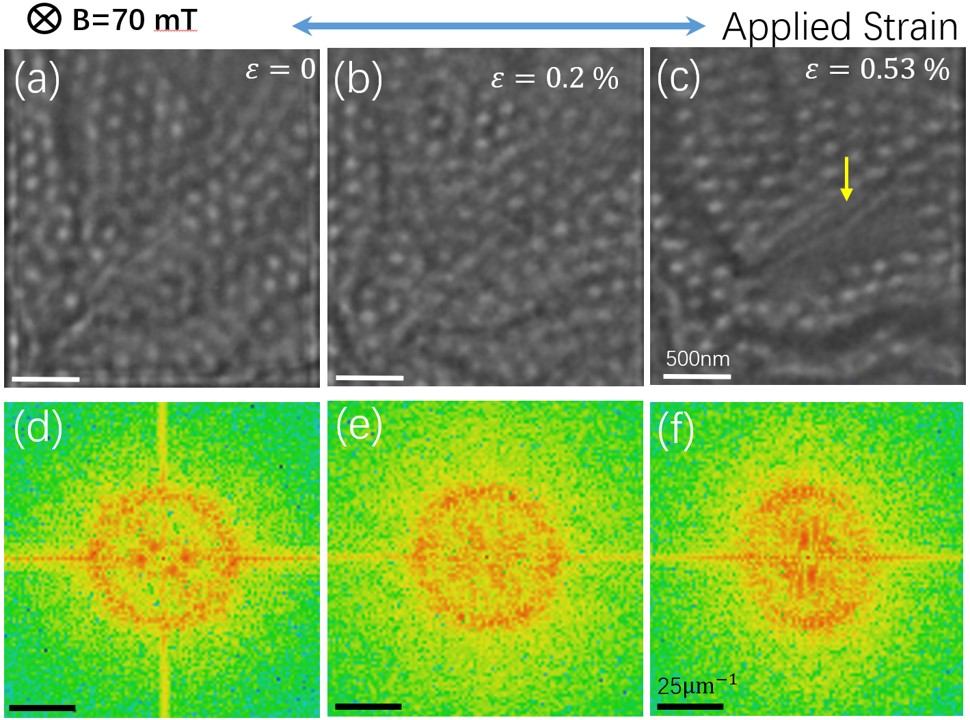} 
    \caption{Magnetic structures and phase diagram in $\sim$190 nm (001) $\text{Co}_8\text{Zn}_{8.5}\text{Mn}_{3.5}$ thin plate with applied strain along the [011] direction. Experiments were performed at room temperature (293 K $\approx 0.87\tc$). (a)-(c) Sequence of LTEM images (taken with an under focus value of $\sim 2$mm) with increasing strain at 70 mT magnetic field. The corresponding Fast Fourier Transform (FFT) analyses are shown in (d)-(f).  Skyrmions are elongated along the strain direction and start to annihilate at a strain of $\varepsilon =0.53\%$. The yellow arrow in panel (c) indicates a conical domain formed through the collapse of the skyrmion lattice.}
    \label{fig_strain_ex}
  \end{figure*}

  \section{IV. Experiment}
\subsection{Experimental Setup}
To validate our theoretical predictions, we experimentally investigated the strain response of skyrmions in a $\text{Co}_8\text{Zn}_{8.5}\text{Mn}_{3.5}$ sample under an applied magnetic field, using a setup nearly identical to the one described in Ref.~\cite{kim2025Topological_}. In-situ tensile testing of a $\sim$190-nm-thick specimen along the [110] direction was performed inside an FEI Tecnai F20 TEM using a Hysitron PI 95 PicoIndenter. The key difference was the application of an external magnetic field to transition the system from its helical ground state into a skyrmion lattice state. LTEM images recorded at a defocus value of
$2$ mm were analyzed quantitatively using phase retrieval based on the  transport of intensity equation \cite{ishizuka2005Phase_}. Comprehensive details regarding sample preparation and stress/strain analysis can be found in the same reference.

\subsection{Experimental Results}
Figure~\ref{fig_EX1} shows the magnetic structures and phase diagram under zero strain. At zero field, the helical phase is stabilized [Fig.~\ref{fig_EX1}(b)]. The measured period gives a DMI length of $L_D \approx 170$ nm. 
Thus, the sample thickness is $L_z = 190/170 L_D \approx 1.1 L_D$, placing it near the transition between thin and thick sample regimes. Under a small applied field ($B \approx 20$ mT), the system partially transforms into the conical phase. The skyrmion lattice phase is stabilized for fields in the range $50 < B < 90$ mT, and a saturated ferromagnetic state is reached at the saturation field $B_D \approx 130$ mT [Fig.~\ref{fig_EX1}(b)]. We chose the skyrmion state at $B = 70$ mT as the initial configuration for strain application, corresponding to $H/H_D = B/B_D = 0.54$, which is close to the field value used in our simulations ($H = 0.55H_D$).

Figure~\ref{fig_strain_ex} presents LTEM observations of skyrmions under applied strain. Two primary effects are observed with increasing strain: circular skyrmions observed at  strains up to $\varepsilon \leqslant 0.2\%$ [Figs.~\ref{fig_strain_ex}(a) and (b)] become elongated at the maximum applied strain of $\varepsilon = 0.53\%$ [Fig.~\ref{fig_strain_ex}(c)], consistent with both our simulation results (Fig.~\ref{fig_mms_inplane}) and previous reports on FeGe thin films \cite{shibataLargeAnisotropicDeformation2015a}. 

Since both skyrmion string tilting and single skyrmion elongation can result in the observed elliptical distortion shown in Fig.~\ref{fig_strain_ex}(c), these two effects cannot be distinguished experimentally. Considering the sample thickness $L_z \approx 1.1 L_D$, the surface twist effect remains significant even in the midsection, leading to partially suppressed tilting. Therefore, elongation likely contributes more significantly to the observed distortion.

Additionally, partial annihilation of skyrmions occurs, and the resulting skyrmion-free regions coalesce into featureless areas
 exhibiting no contrast [Figs.~\ref{fig_strain_ex}(c)]. 
 This behavior is in close agreement with our simulations [Fig.~\ref{fig_mms_inplane}(c)]. The observed anti-cluster formation serves as an indicator of conical phase formation. 
As established in our simulations, this collapse pathway differs fundamentally from field- or anisotropy-induced annihilation mechanisms. 
This clear distinction provides strong evidence that, in this material, the dominant effect of strain is to modulate the DMI, rather than to induce magnetic anisotropy.

\section{V. Conclusions}

In summary, we have established a comprehensive framework for strain control of magnetic skyrmions through integrated analytical modeling, micromagnetic simulations, and \textit{in situ} Lorentz transmission electron microscopy (LTEM)experiments. Our analytical model reveals that strain induces opposing-direction tilting of skyrmion strings in the midsection of thick samples, where surface twist effects are negligible. Numerical simulations confirm this model and further demonstrate that surface twist suppresses tilting in thin films and near surface regions of thick samples. Simulations also reveal that strain drives the system from fragmented multi-domain configurations at low strains toward consolidated single-domain states at high strains. The tilt geometry weakens inter-layer coupling, enabling strain to rupture skyrmion strings at their midsections, generating bobber pairs that subsequently shrink and annihilate at the surfaces, thereby erasing individual skyrmion strings.

In thick samples ($L_z > L_D$), the collapse of the entire skyrmion lattice proceeds through temperature-modulated phase transitions—exhibiting first-order behavior at low temperatures and transitioning to second-order near $T_c$. Reducing the sample thickness significantly affects the critical strains: while the onset strain $\eta_c^{\text{onset}}$ for skyrmion annihilation remains largely thickness-independent, 
the completion strain $\eta_c^{\text{finish}}$ increases markedly in thin samples ($L_z < L_D$) due to tilt suppression.

Experimental validation in $\text{Co}_8\text{Zn}_{8.5}\text{Mn}_{3.5}$ reveals a sequential response to strain: initial axis-aligned elongation below a threshold, followed by dissolution into conical domains via anti-cluster formation.
This pathway is distinct from field-induced annihilation, which exhibits gradual dilution into a ferromagnetic background. These findings provide an alternative strategy for manipulating topological spin textures and offer valuable insights for developing strain-engineered spintronic devices.

 \begin{acknowledgments}

   H.Z. is supported in part by National Natural Science Foundation of China (Grant No. 11704067).
  H.Z. thanks the computational resources from
  the Southeast University Campus-Wide Computing Platform.
  This work is supported in part by Laboratory Directed Research and Development funds through Ames National Laboratory (H.Z., T.K., L.Z.). L.K. was supported by the U.S. Department of Energy, Office of Science, Office of Basic Energy Sciences, Materials Sciences and Engineering Division, Early Career Research Program following conception and initial work supported by LDRD. Ames National Laboratory is operated for the U.S. Department of Energy by Iowa State University under Contract No. DE-AC02-07CH11358.  All TEM and related work were performed using instruments in the Sensitive Instrument Facility in Ames National Lab. Current address of Tae-Hoon Kim: Department of Materials Science and Engineering, Chonnam National University, Gwangju 61186, Republic of Korea.
\end{acknowledgments}

\end{document}